November 13, 2013

# Status and prospects for BESIII

Zhong-Hua Qin[1]
ON BEHALF OF THE BESIII COLLABORATION

*Institute of High Energy Physics*
*Chinese Academy of Science, 100049, Beijing, China*

In this paper the status and prospects for BESIII is presented. The BEPCII machine is in good status and recently reached its highest luminosity of $7.08\times10^{32}$ $cm^{-2}$ $s^{-1}$. The BESIII detectors have been running stably since 2009. Upgrade plans for its sub-detectors have been proposed and the upgrade is undergoing. The major upgrade comes from the drift chamber and the time of flight counter. The other sub-detectors show good status and no upgrade are needed. Given the current status and upgrade plan of the BESIII detectors, it's foreseen that the BESIII is able to work for 8~10 years from now.

---

[1]E-mail:qinzh@ihep.ac.cn

# 1 Introduction and status of BEPCII

The tau-charm collider BEPCII [1] is a high luminosity, double-ring $e^+ e^-$ collider as shown in Fig.1 and Fig.2. The luminosity designed for BEPCII is 100 times higher than BEPC by increasing the number of bunch in storage ring and reducing β function in interaction region. The main parameters for BEPCII are shown in Table 1.

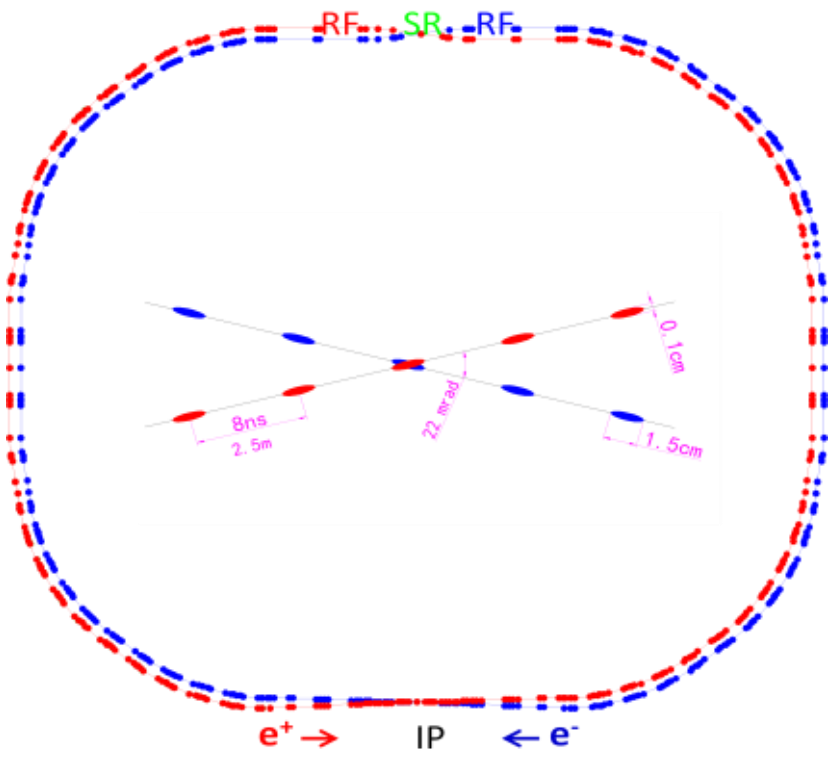

Figure 1: The schematic view of BEPCII

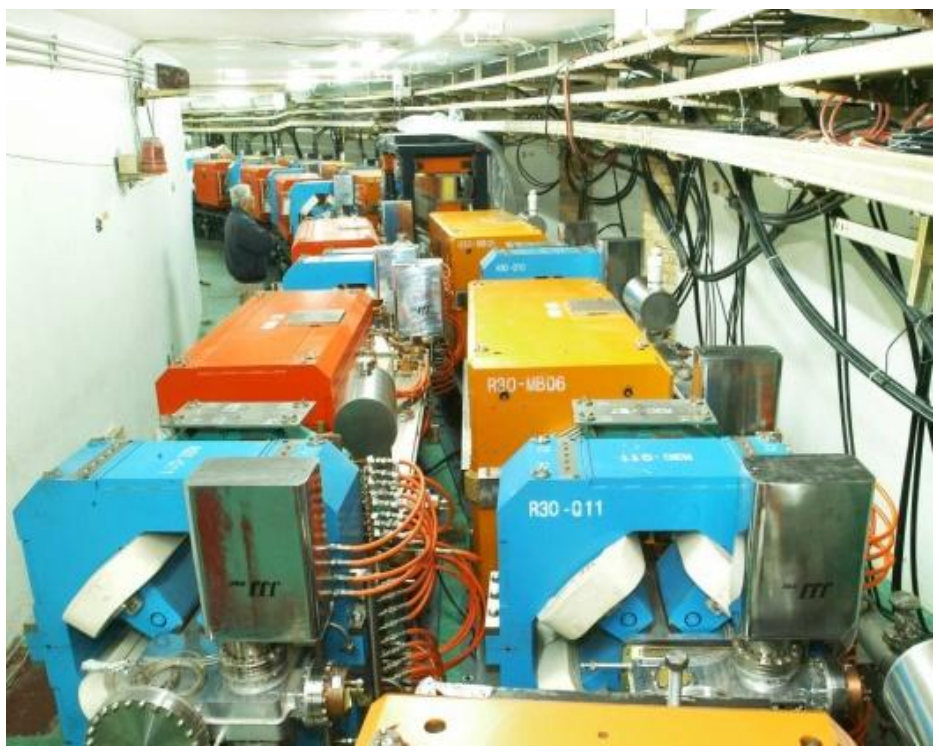

Figure 2: A picture of the double rings

| | |
|---|---|
| Beam energy | 1.0-2.3 GeV |
| Design Luminosity | $1\times10^{33}$ $cm^{-2}s^{-1}$ |
| Optimum energy | 1.89 GeV |
| Energy spread | $5.16\times10^{-4}$ |
| No. of bunches | 93 |
| Small □ *y** | 1.5 cm |
| Total current | 0.91 A |
| SR mode | 0.25A @ 2.5 GeV |
| Large cross angle | 22 mrad |

Table 1: Main parameters of BEPCII

□

BEPCII started its running for data taking from 2009, the luminosity was increasing gradually. During 2013, a record peak luminosity of $7.08\times10^{32}$ $cm^{-2}$ $s^{-1}$ has been reached in a period of machine study as shown in Fig.3. The new daily record lumi. of $17pb^{-1}$ and weekly of $169pb^{-1}$ ,have also been obtained during 2013 year's data taking, as detailed in Fig.4.

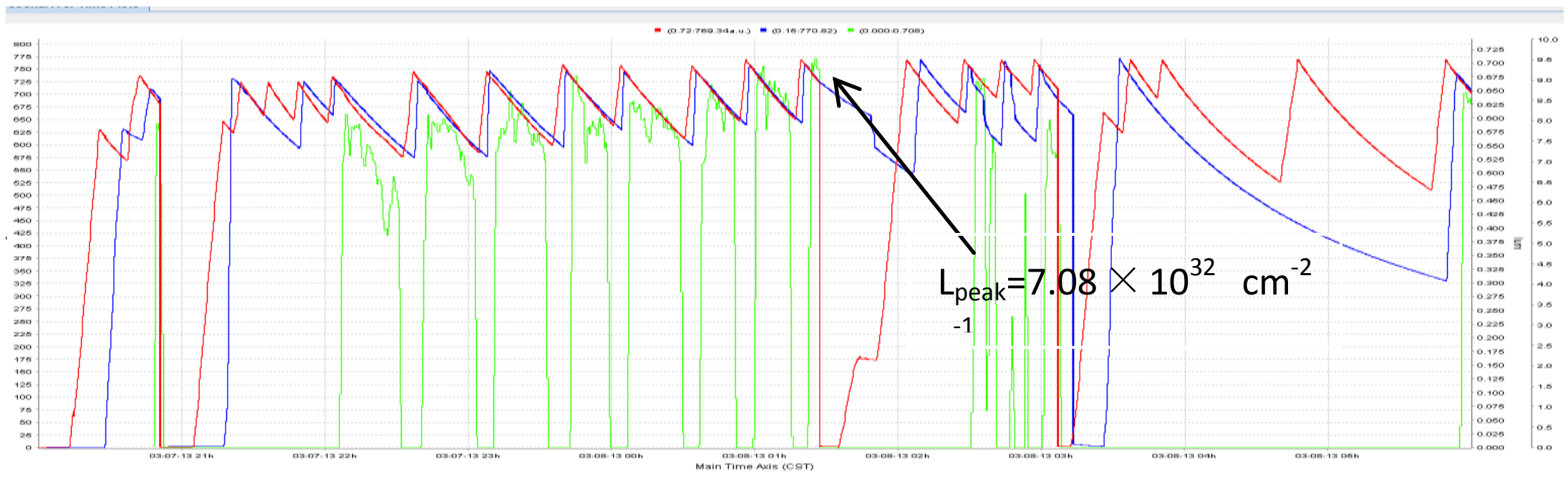

Figure 3: A new record lumi. of $7.08\times10^{32}$ $cm^{-2}$ in 2013

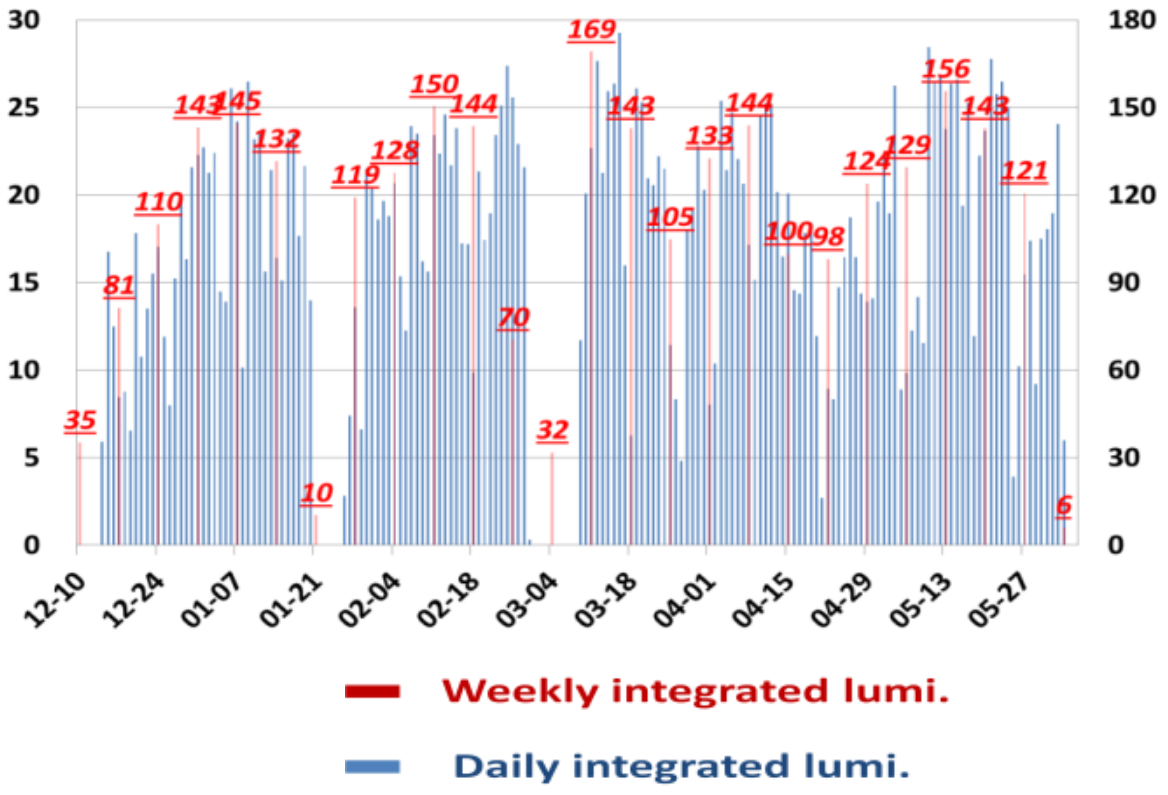

Figure 4: Lumi. histogram for 2013 data taking.

Totally 2.9fb$^{-1}$ data have been collected in 2013 mainly for Y (4040), Y(4260) and Y (4360). The machine is planning to run in the high limit of its energy range(~4.6GeV) for next round of data taking so as to explore the new physics.

## 2 Introduction of BESIII

BESIII [2] [3] is a high performance spectrometer designed to work with high lumi. machine. It consists of five sub-detectors: the main drift chamber (MDC), the time of flight counter (TOF), the electromagnetic calorimeter (EMC), the super-conducting solenoid magnet (SSM) and the muon chamber (MUC). A schematic view together with a brief induction of the BESIII is shown in Fig. 5.

The BESIII officially started data taking in 2009 and running smoothly over the past 4 years. A detailed status and upgrade of BESIII is introduced in the later sessions.

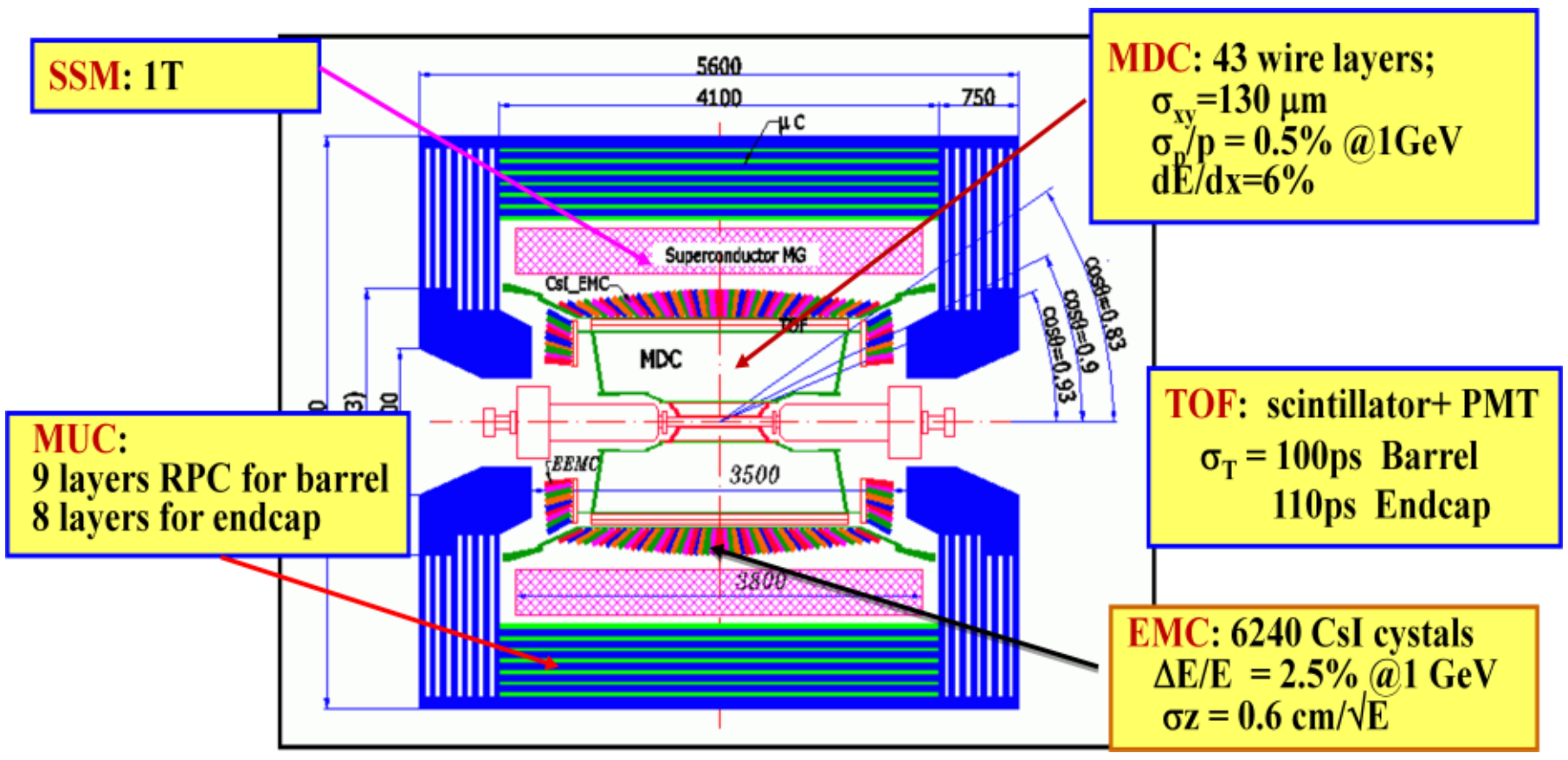

Figure 5: A schematic view and brief introduction of BESIII detector

## 3 Status and upgrade of MDC

### a) the status of MDC

MDC finished its construction and cosmic-ray test in 2007, the chamber was installed to its final position and commissioned in 2008 and started its running for data taking from 2009. Currently the chamber has reached its best performance as shown in Fig.6, and the performance values are summarized in Table 2.

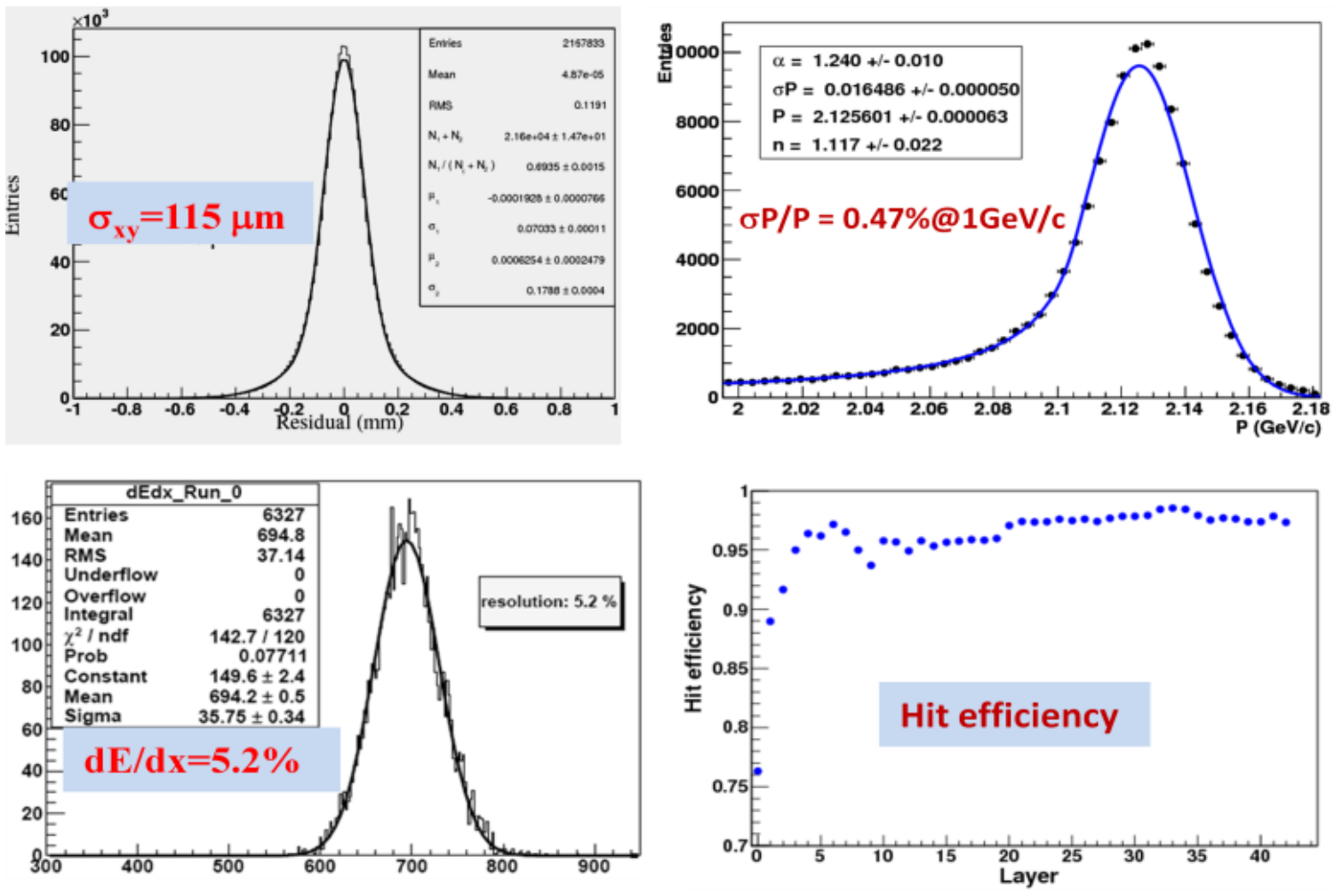

Figure 6: Figures for MDC performance

| | Achieved value | Design value |
|---|---|---|
| Spatial resolution $\sigma_{xy}$ | 115 μm | 130μm |
| Momentum resolution $\sigma_p/p$ | 0.47% @1GeV | 0.5% @1GeV |
| Energy loss resolution dE/dx | 5.2% | 6~7% |
| Hit efficiency ε | ≥95% | 98% |

Table 2: Summary of MDC performance values

The drift chamber currently is still in good quality, no wire is broken for totally 28000 wires. There are 22 dead channels, and most of them are due to a problem with the front-end electronics. Since the number of dead channels is only ~0.3% of the total channels, there is no influence on tracking efficiency.

However, the chamber has been suffering from a heavy beam-related background from accelerator since the very beginning. The high voltages for the first 4 layers of the chamber are 96%, 97%, 98%, and 99% of the nominal values, respectively. Also, due to the high counting rate caused by the background, the inner chamber appeared self-sustained discharging in 2012, the current of some sense wires increase dramatically after a few minutes of running, as one can see from Fig.7. This problem was finally identified as Malter effect [4] which is a typical aging for wire chamber due to a thin insulating film (some polymers produced during gas ionization process) formed on the surface of cathode wire. To solve this problem, ~2000ppm water vapor was added into the chamber gases then the discharge disappeared.

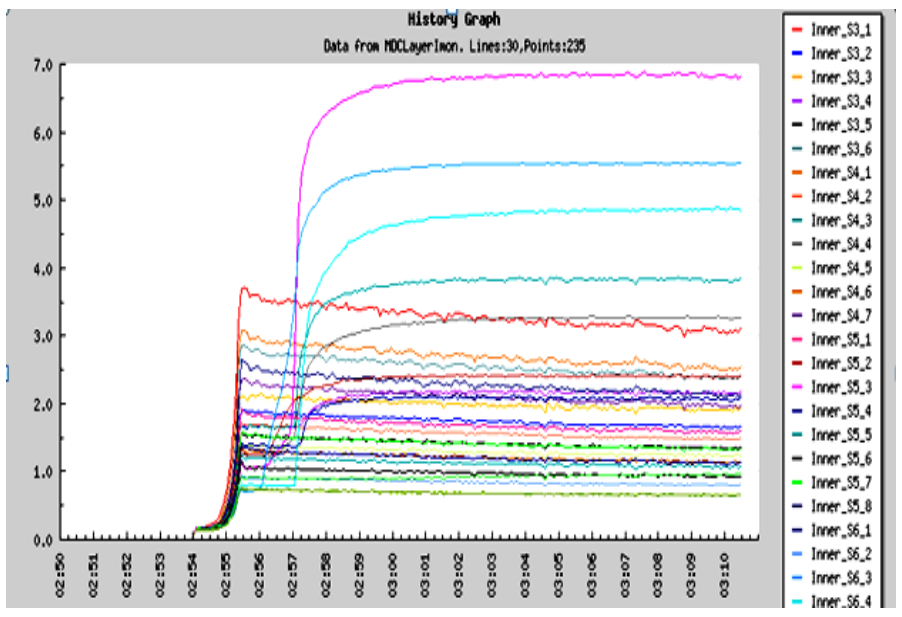

Figure 7: Malter effect of the chamber

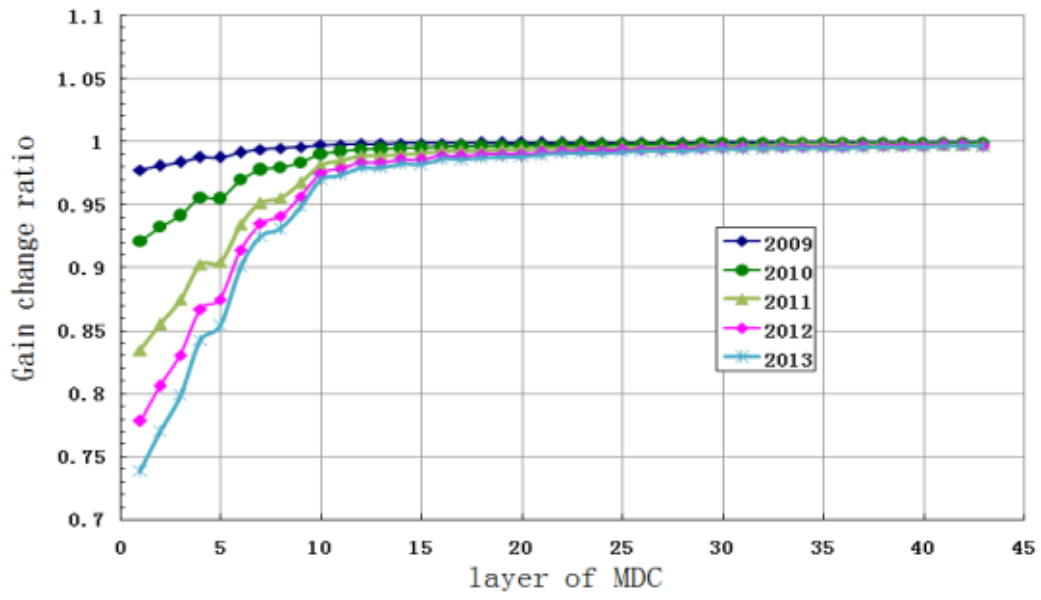

Figure 8: Gain drop of the chamber

The chamber is also observed gain drop for sense wire as shown in Fig.8. Compared to 2009, the gas gain of first 5 layers decreased about 26% -14% , and the gain of the first 10 layers also show an obvious decrease. But for the outer chamber, the drop is negligible.

### b) The upgrade of MDC

Given the aging mentioned above, upgrade of the inner drift chamber is necessary:

A Short-term upgrade plan: build a new inner drift chamber to replace the current one in summer, 2015. This project was already approved by CAS with a 3.6M rmb budget;

A Long-term upgrade plan: a 3-layer CGEMs inner tracker will be built by 2017, and the chamber could be replaced in 2017 or later in 2018. The funding for R&D was also approved with totally 360 k€ and co-funded by Itlay and China.

## i. the new inner drift chamber

A new inner drift chamber is under construction for the 2015 upgrade, it also functions as a backup chamber in case serious problem appearing for the current one. A feature of the new chamber is that the length of wires is shortened , to reduce the counting rate (currently it's $2.2{\times}10^5$Hz for heaviest background) while keeping same acceptance as current one. The comparison of the two chambers is shown in Fig.9 (a) and Fig.9 (b). It's foreseen that the counting rate can be reduced by 30% for the inner most layer.

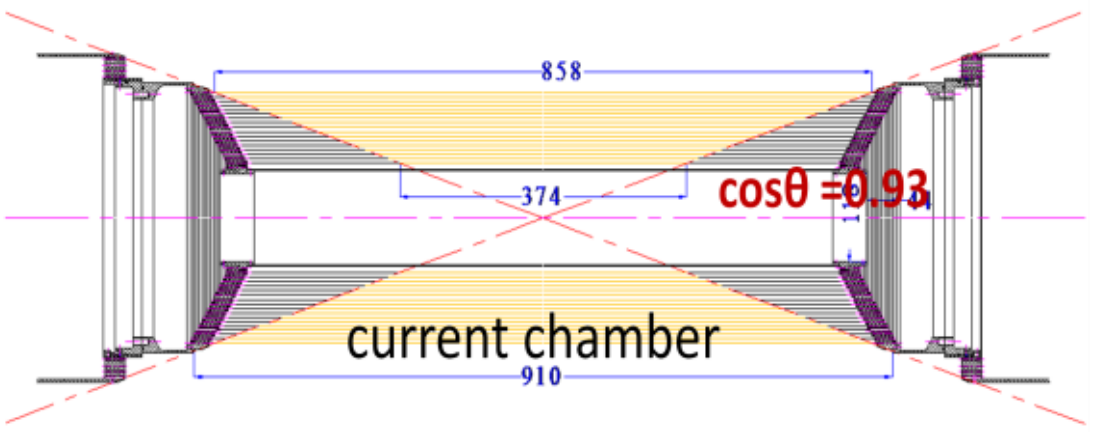

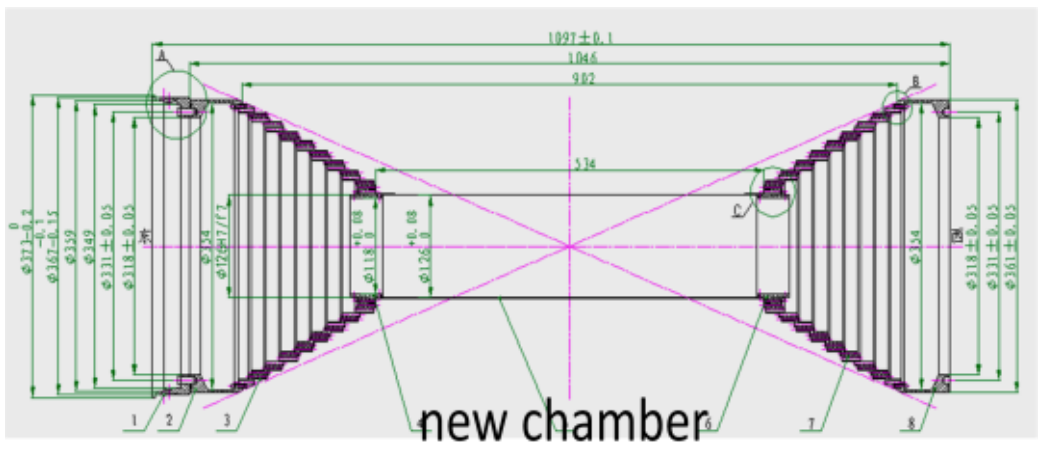

Figure 9(a): the current chamber (b): the new chamber

Simulation shows that the new chamber has similar momentum resolution and tracking efficiency compared to the current one, and better Z resolution, as shown in Fig. 10(a), (b), (c).

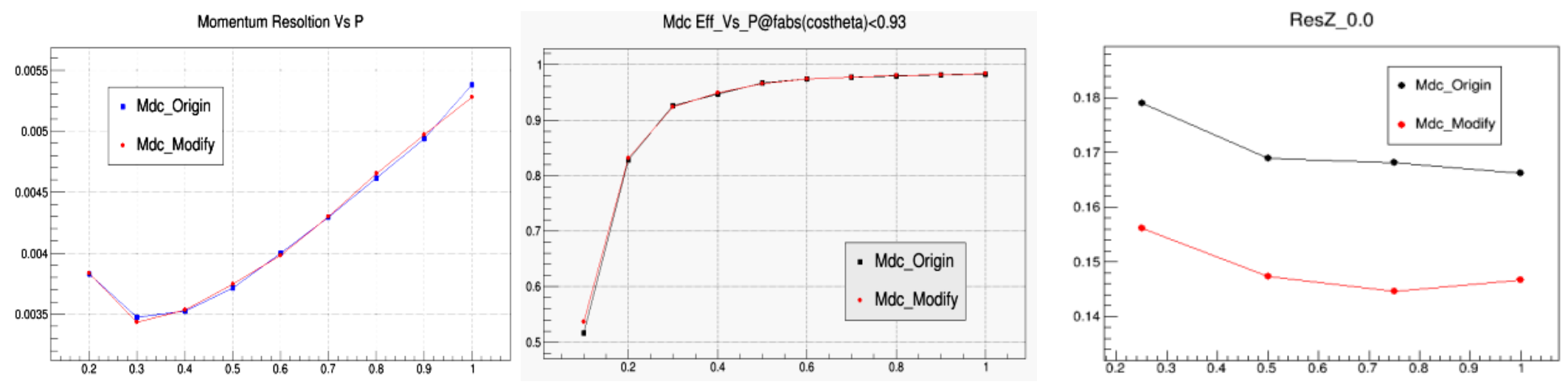

Figure 10: (a) Momentum resolution. (b): Tracking efficiency. (c): dZ.

### ii. The CGEM inner tracker

The Cylindrical GEM (CGEM) option is proposed by the BESIII Italy group, which is a 3-layer CGEM foils worked as an KlOE2-like inner tracker. The conceptual view of the CGEM is shown in Fig.11 and a possible layout of the inner tracker is shown in Fig.12.

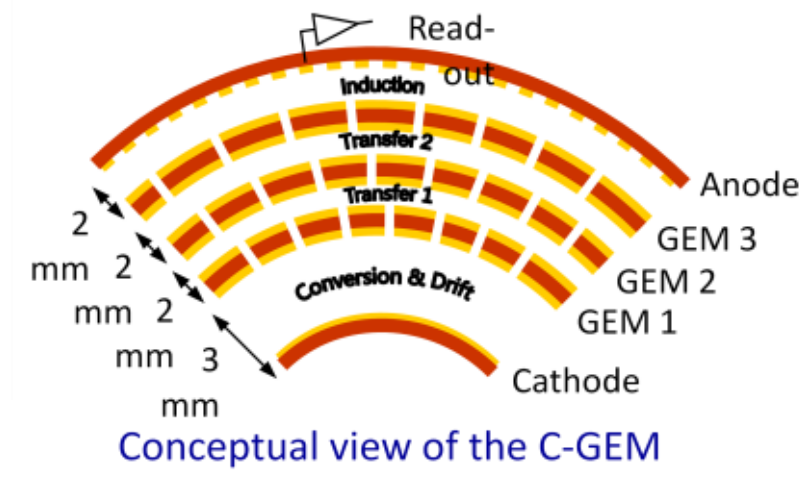

Figure11: Conceptual view of CGEM

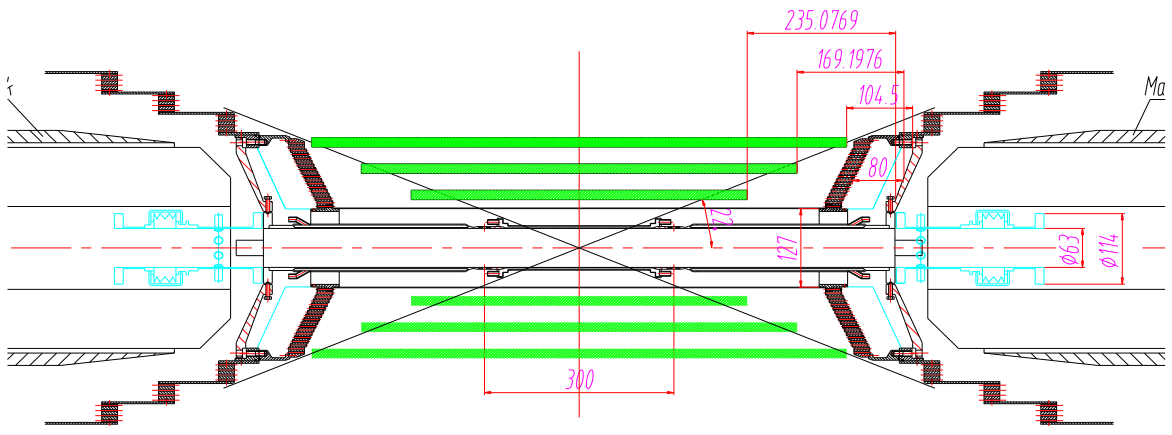

Figure12: CGEM inner tracker in MDC

Simulation shows comparable momentum resolution and R resolution, and improved Z resolution than the current chamber, as shown in Fig. 13 (a),(b),(c)

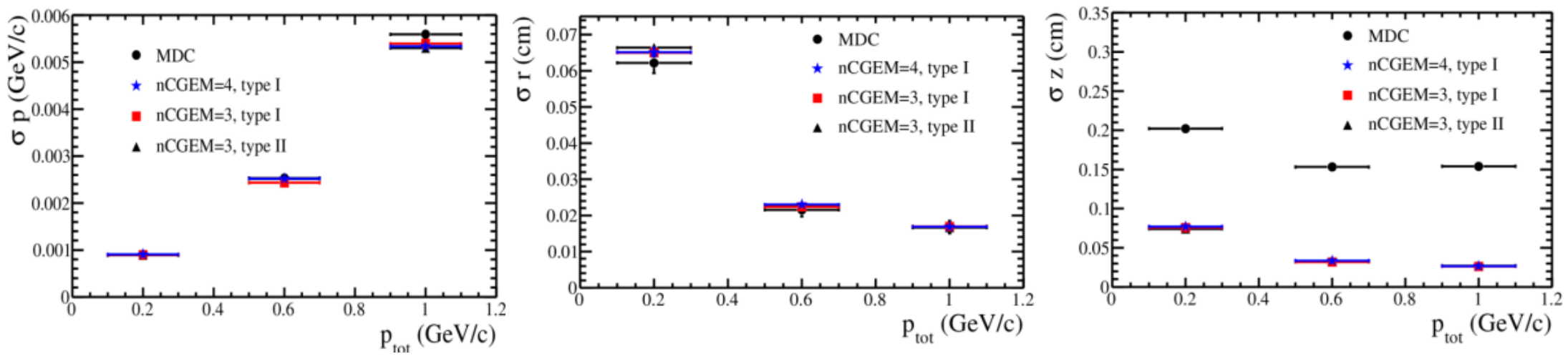

Figure 13(a): Momentum resolution (b): R resolution (c): Z resolution

## 4 Status and upgrade of TOF

### a) The status of TOF

BESIII TOF is divided into two parts: one barrel TOF (BTOF) and two endcap TOF(ETOF). In which barrel TOF is two layer scintillators with double-readout and 88 blocks for each layer, and the endcap TOF is one layer with single-readout and 48 for each part. The schematic view of TOF is shown in Fig. 14.

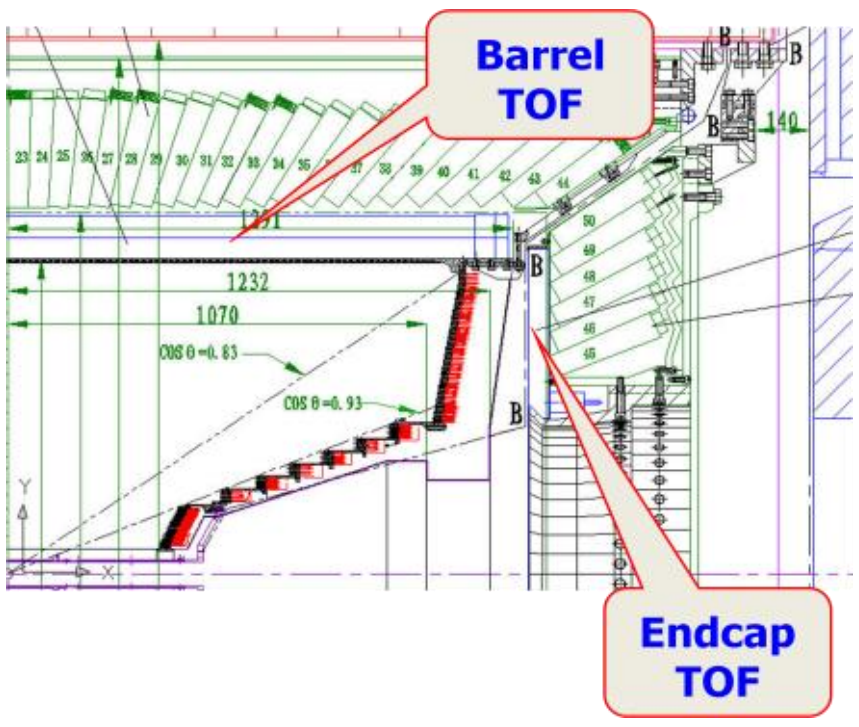

Figure 14: A schematic view of BTOF and ETOF

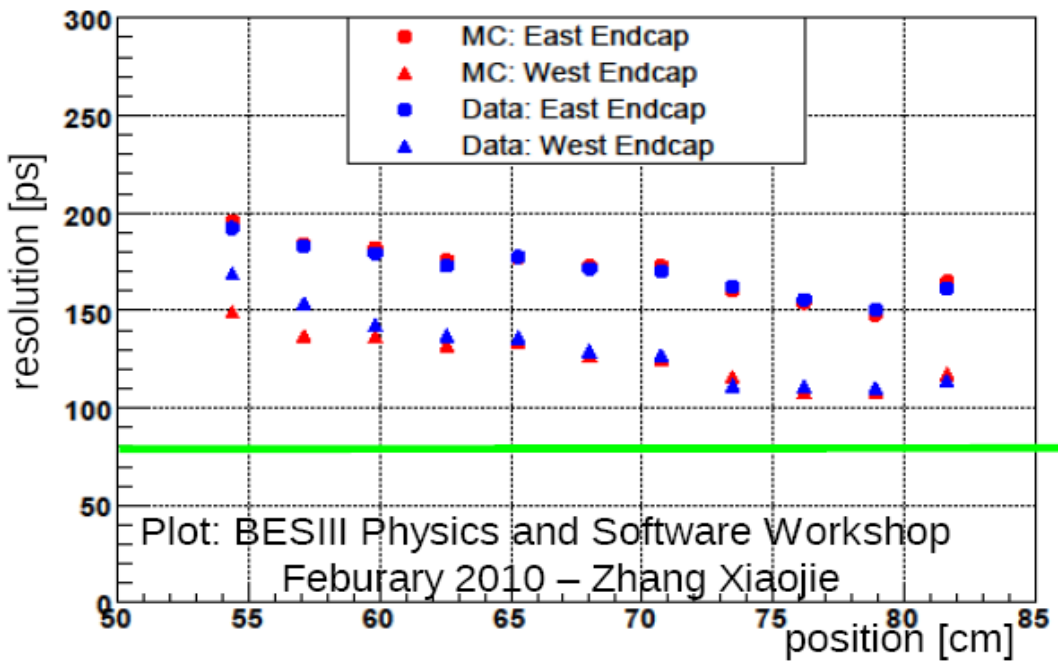

Figure 15: time resolution of ETOF

For BEOF, the time resolution is better than 70ps and efficiency larger than 96% for most of the years. For ETOF, the performance is more affected by noise and background. The time resolution for Bhabha events varies from 100ps~200ps, as shown in Fig. 15.

There is some aging effect observed for TOF, shown as Fig. 22(a),(b) for attenuation length and Fig.23(a),(b) for light yield, respectively. As seen from Fig.22, the average decreasing rate of attenuation length is 4.3% per year and the efficiency decreases with the attenuation length decreasing. The change of time resolution is not observed.

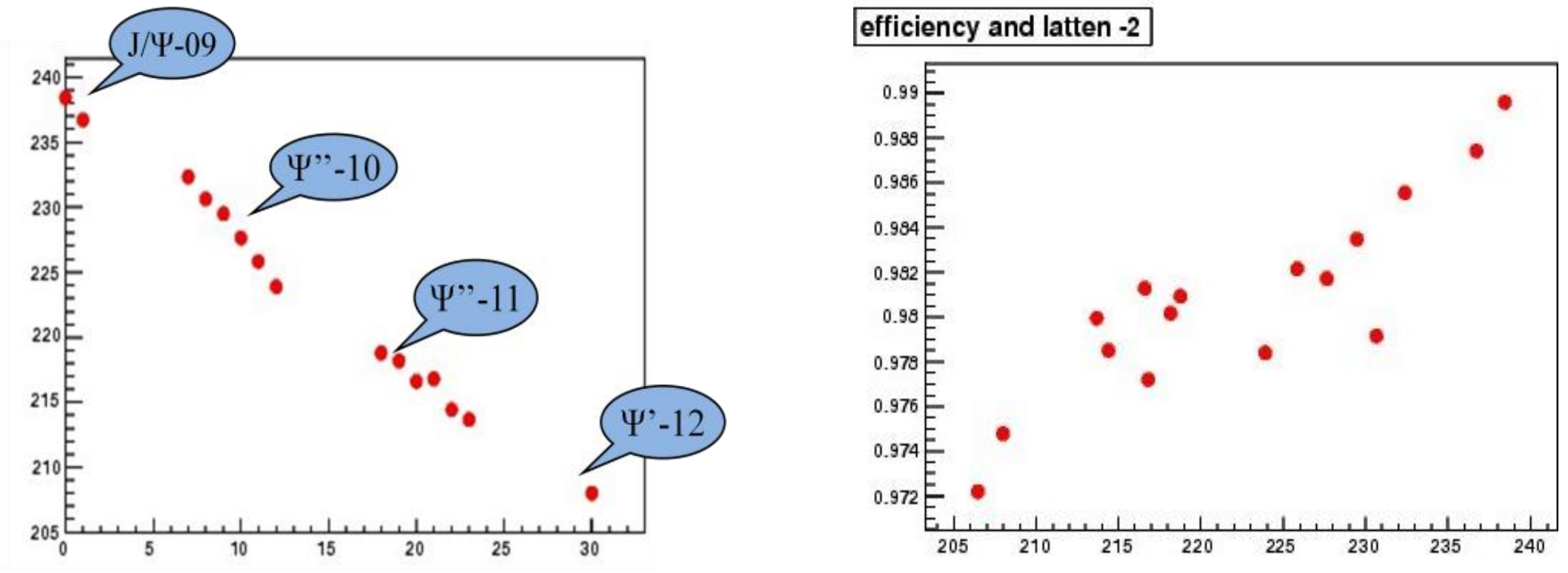

Figure22(a):Attenuation length vs time(/month) (b): Attenuation length vs Efficiency

For light yield, as seen from Fig. 23(a), the average decreasing rate of light yield is 0.49% per year. So the QTC is needed to be corrected for different time point.

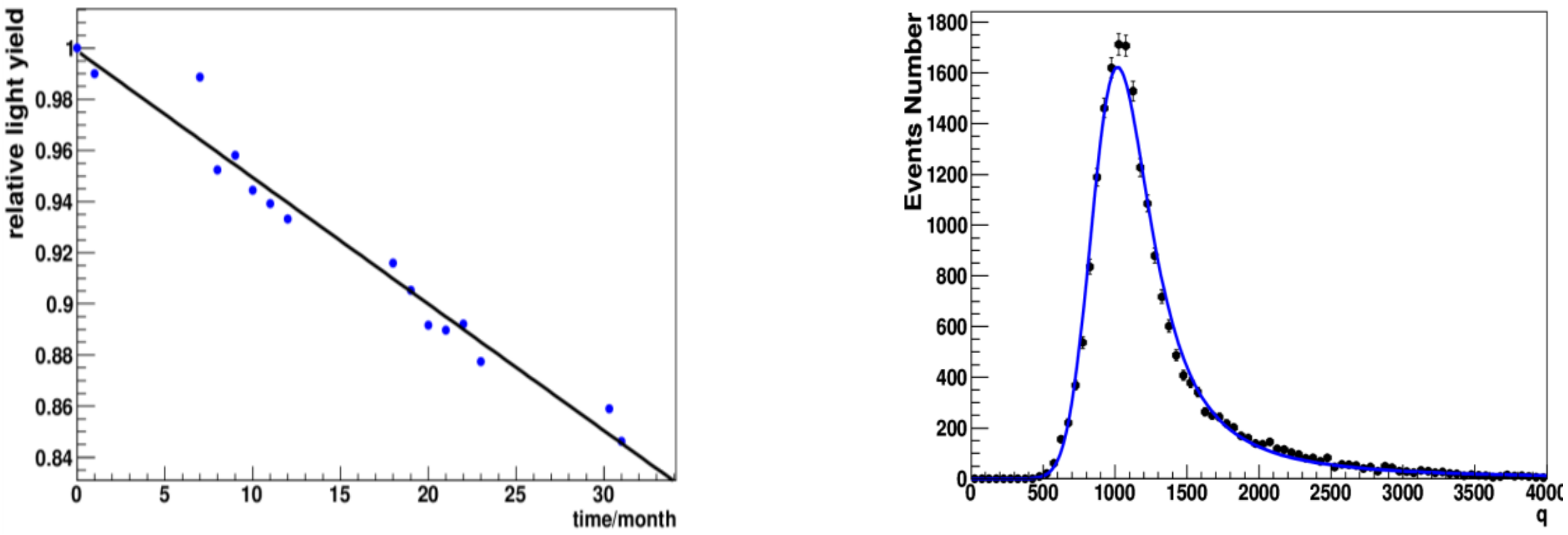

Figure 23 (a): MPV of QTC vs time (b): Corrected QTC

### b) Upgrade of ETOF

The current BESIII ETOF is scintillator plus PMT technology, the time resolution for μ is 110ps, for electron 148ps and for pion 138ps. On average, for 2σ π/K separation power the energy is about 1.0GeV, due to multiple scattering from the material, tracking extrapolation issue and multi-hits produced. To achieve accurate measurements in the T-charm energy region and search for rare decays (for example $D^0D^0$bar Mixing and so on), we need good PID for hadrons up to 1.4GeV for 2σ π/K separation power, as shown in Fig. 24.

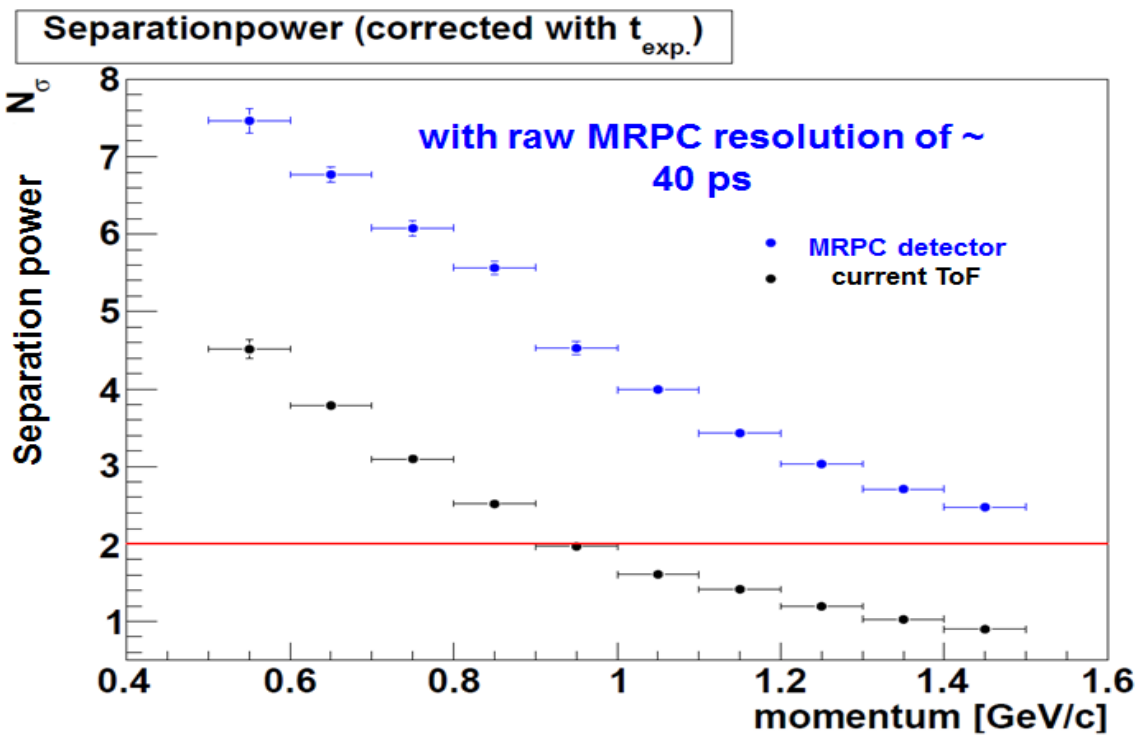

Figure 24: separation power of current and upgraded ETOF

A Upgrade ETOF with MRPC can be less affected by scattering and tracking with more readout pads, the total resolution can be less than 80ps (MRPC: 50ps; Non-intrinsic: 50ps)

The design for ETOF upgrade includes two ETOF rings. For each ring, there are 36 overlapping MRPC modules, which are sealed in gas-tight boxes with a thickness of each box less than 25 mm, as shown in Fig. 25. The FEE boards are mounted in between the two adjacent boxes to save space, as shown in Fig.26

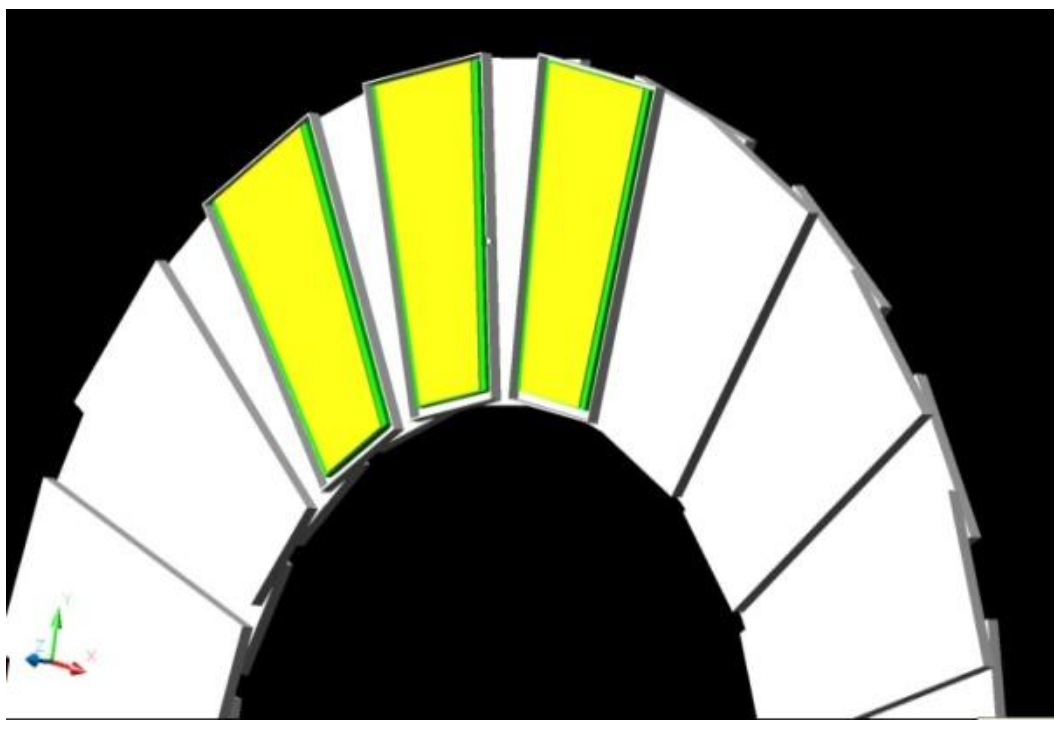
Figure 25: MRPC module

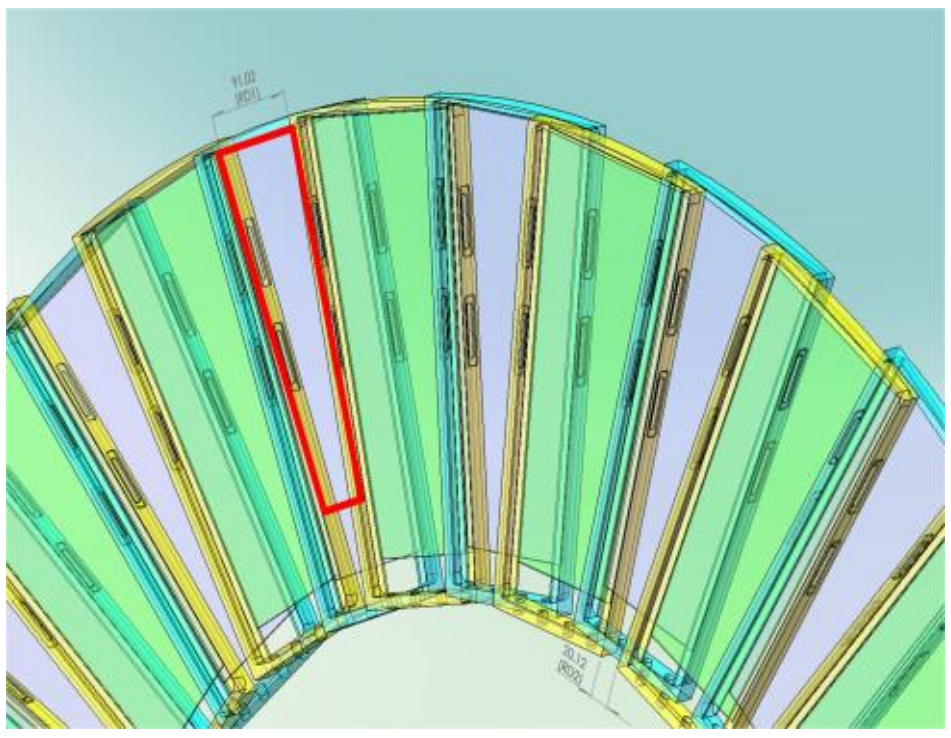
Figure 26: FEE boards

Structure of the MRPC is shown in Fig. 27. It is composed of a stack of resistive plates with 12 gas gaps. The resistive plate is made from floating glass (~$10^{12}$–$10^{13}$ Ω*cm), and with 14KV high voltage applied. The total thickness of a MRPC module is ~20 mm. The readout strip is shown in Fig.28, it's a double sided read-out with a width of 2.5 cm and length of 8.6-14.1cm. There are 24 channels for each module and totally 24 x 36 x 2 = 1728 channels.

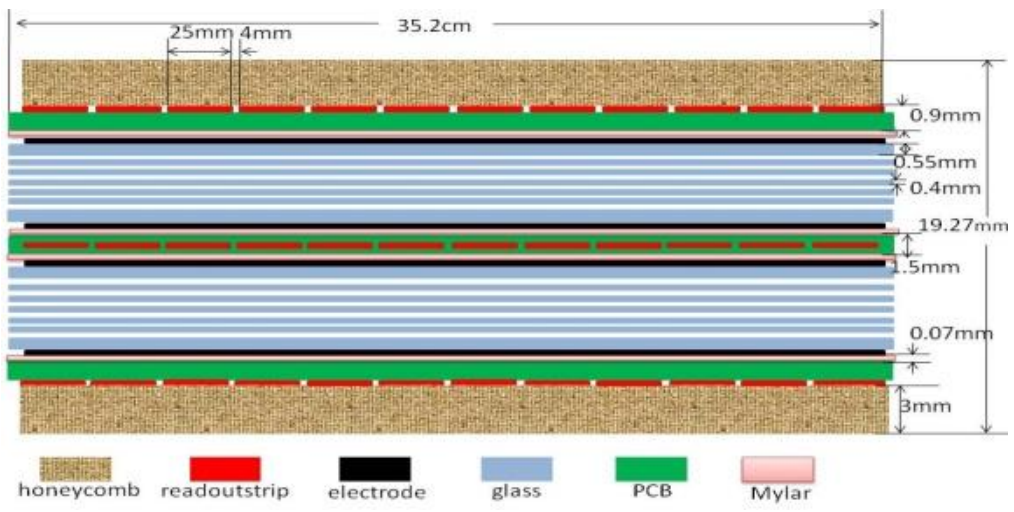

Figure 27: MRPC Layout

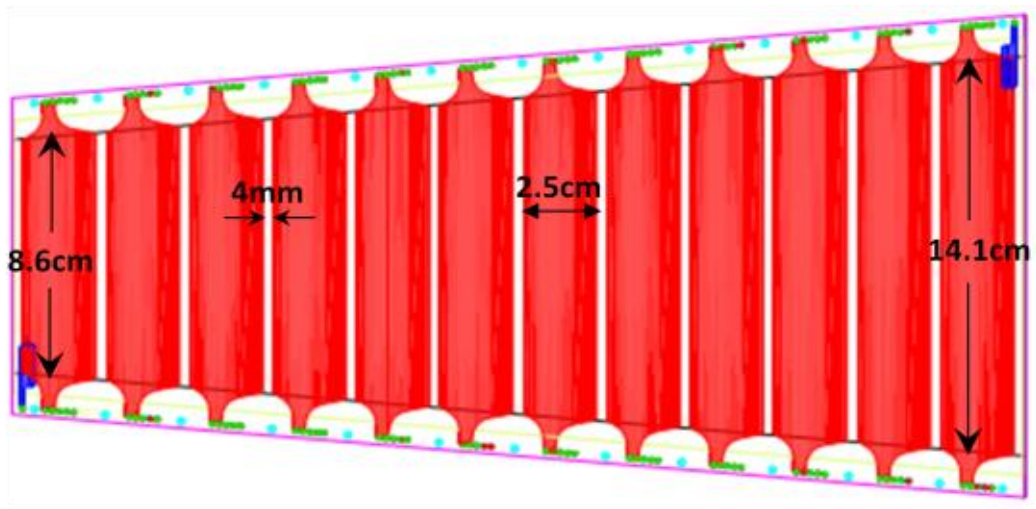

Figure 28: MRPC readout strip

Upgrade of EFOF is planned to be finished in 2015, and the replacement is scheduled in summer of 2015.

## 5 The status of EMC

The performance of EMC is very good through all the past years and there is no dead channel until now. Fig.29 shows the energy resolution of EMC, one can see most crystals keep unchanged compared to 2009. And for light output, it is also not decreased distinctly as shown in Fig.30. There are only 32 out of 6240 (~0.5%) crystals decrease by 15%, and 12 out of 6240 (~0.2%) decrease by 20%;

No upgrade foreseen for EMC in near future given its perfect performance.

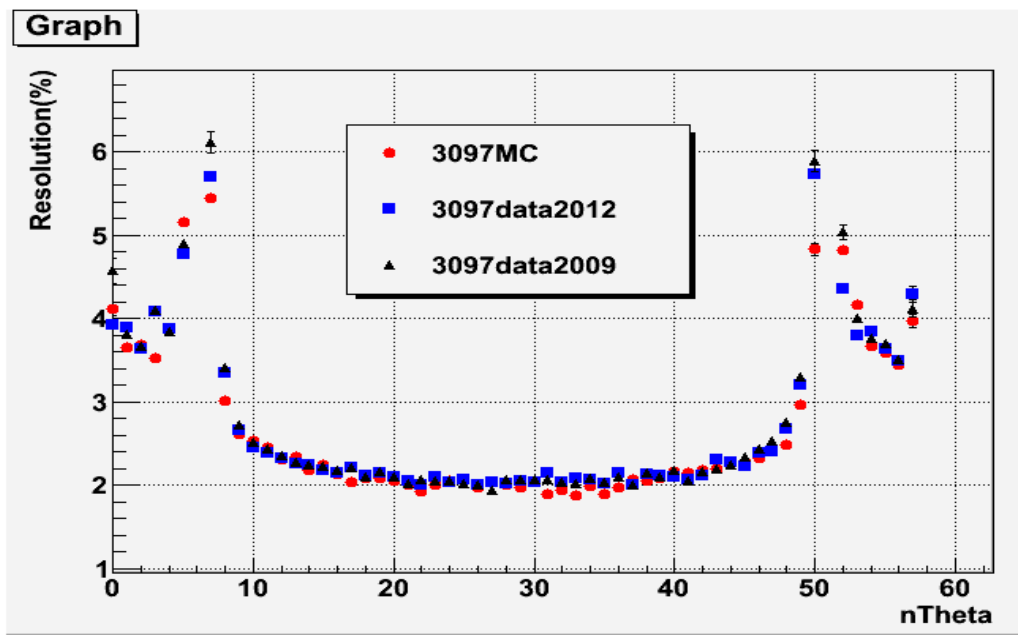

Figure 29: Energy resolution

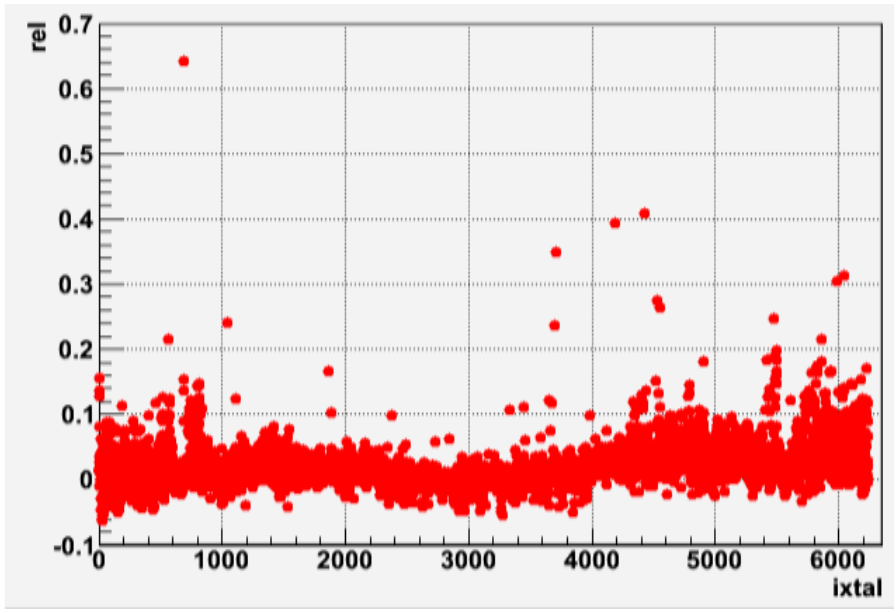

Figure 30: Relative decrease of light output

## 6 Status of SSM

BESIII SSM Worked stably on 3369A providing 1.0Tesla magnetic field since 2009. Sometimes it's quenched due to the cryogenic system failure. In Feb. 2012, temperature of the current leads transition section in valve box rose obviously from 6.30K to 6.38K, as shown in Fig.31, for safety reason SSM was shifted to 0.9T for later running of 2012.

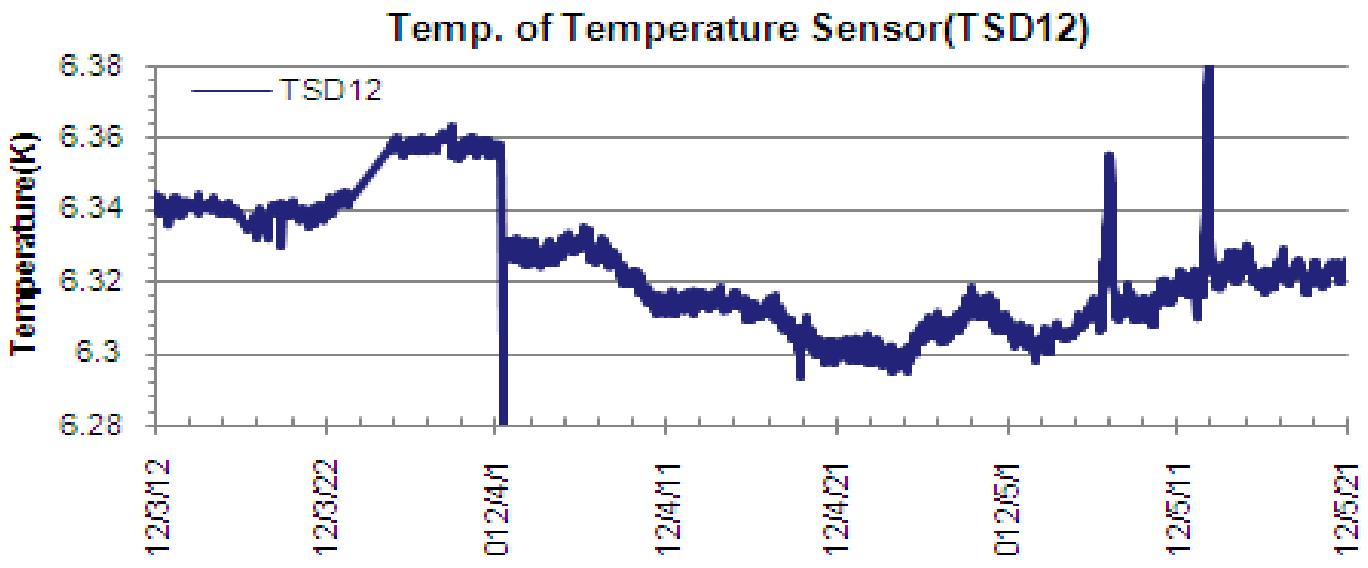

Figure 31: Temperature of current leads transition section

Tests showed that SSM is actually able to work under higher temperature of ~6.5K. The magnetic field was set back to 1.0T from 2013 data taking, and SSM worked well since then. Construction of a backup valve box is under consideration.

## 7 Status of MUC

Major upgrade has already been done during summers of 2011 and 2012 for MUC

front-end readout electronics (FECs), due to the failure in a comparator chip. After then, the chamber worked stably.

In 2013, High voltage of MUC was decreased from 7.7kV to 7.5kV, to protect both the chamber and its electronic, but without deteriorating its performance. As shown in Fig. 32(a),(b), the efficiency is still reaching 95% at 7.5kV, and there is almost no affect on spatial resolution and still is 26mm after reducing HV. The noise level is less than 0.01Hz/cm2, thus no influence on tracking.

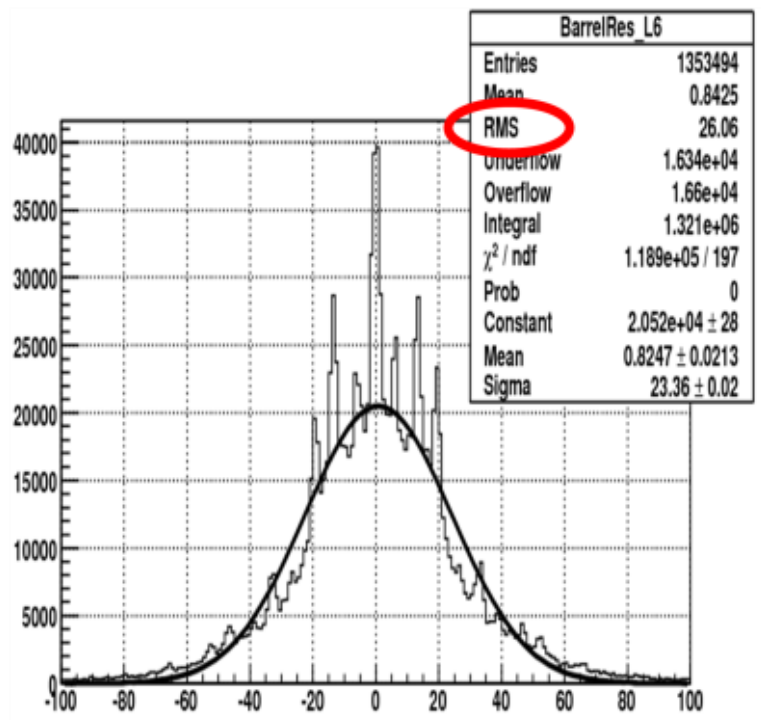

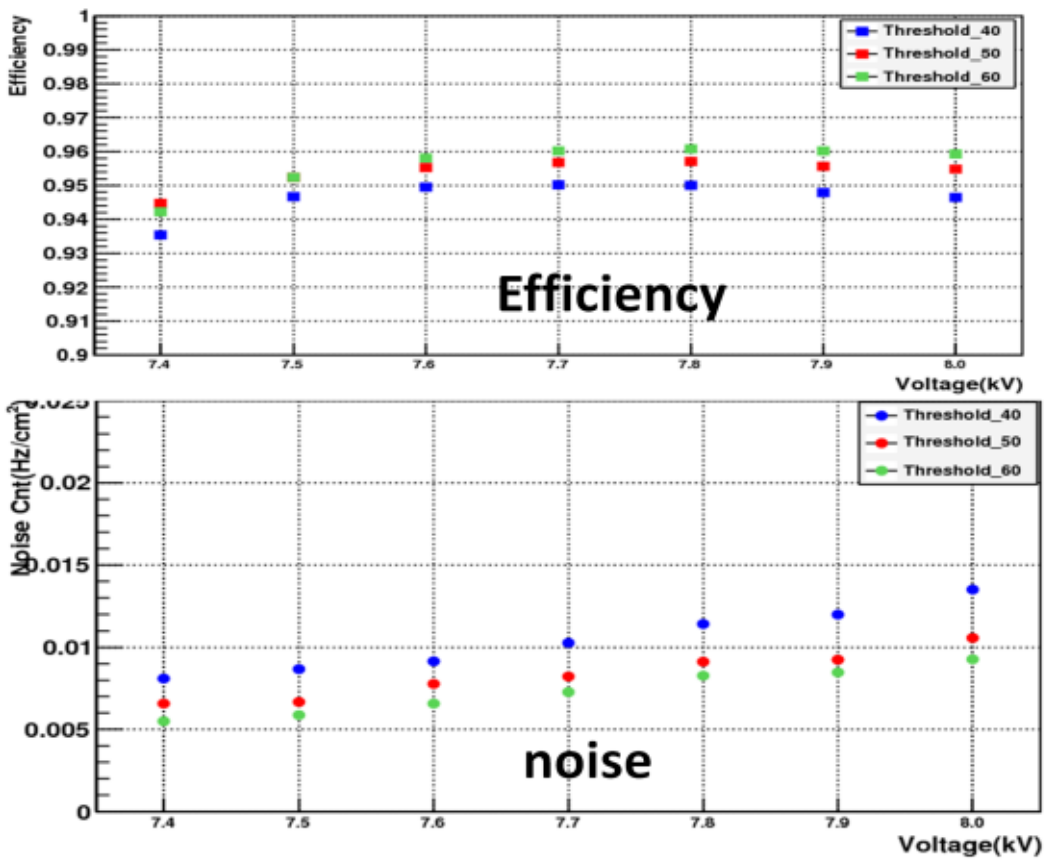

Figure 32(a): spatial resolution (b): Efficiency and noise level

## 8 Summary

An overview of the BESIII sub-detectors have been given, and currently BESIII works well with high efficiency. MDC has already started a short-term upgrade with a new inner drift chamber, the installation is scheduled in summer of 2015. The long-term upgrade plan with CGEM is supposed to be finished in 2017 or 2018; Currently both upgrade activities are undergoing. An upgrade of TOF with MRPC is also started and will be finished in summer of 2015, the same time scheduled for new inner drift chamber.For others sub-detectors (EMC, SSM, MUC), no major upgrade are needed in near future, and longer-term upgrade will start to consider. Given the current status, the BESIII detectors can still work for 8 to10 years.

## ACKNOWLEDGEMENTS

I would like to thank the organizers for the successful conference. And thank the colleagues in BEPCII and BESIII for providing the materials.